# Performance and Fault Tolerance in the StoreTorrent Parallel Filesystem

Federico D. Sacerdoti
NY NY, 10128
USA

## Abstract

With a goal of supporting the timely and cost-effective analysis of Terabyte datasets on commodity components, we present and evaluate StoreTorrent, a simple distributed filesystem with integrated fault tolerance for efficient handling of small data records. Our contributions include an application-OS pipelining technique and metadata structure to increase small write and read performance by a factor of 1-10, and the use of peer-to-peer communication of replica-location indexes to avoid transferring data during parallel analysis even in a degraded state. We evaluated StoreTorrent, PVFS, and Gluster filesystems using 70 storage nodes and 560 parallel clients on an 8-core/node Ethernet cluster with directly attached SATA disks. StoreTorrent performed parallel small writes at an aggregate rate of 1.69 GB/s, and supported reads over the network at 8.47 GB/s. We ported a parallel analysis task and demonstrate it achieved parallel reads at the full aggregate speed of the storage node local filesystems.

## 1. Introduction

Scientific digital instruments, computational simulations of physical systems, and other digital sources have created the need for large scale storage systems, with capacity on the order of Petabytes per year. Increasingly, simple or extensive processing of this raw data is needed to gain final results and arrive at essential discovery, pushing the limits of distributed filesystems.

While some data can be written in large chunks and is always accessed sequentially during analysis, other data naturally lends itself to small records. Analysis of this data may require efficient random access to a particular record, for example the case of analyzing a stream made up of every Nth record in the dataset, or every record named by a specific symbol. This work addresses shortcomings of existing distributed filesystems with regard to 1-2MB writes, and supports a small data record of this size as the minimum addressable unit for high-speed random access during analysis.

We target an HPC scientific environment that contains a commodity cluster connected by a local area network, whose nodes have several inexpensive, high capacity storage drives. Additionally the environment contains a resource manager that can start parallel jobs on the cluster, and a reliable network fileserver providing user data to each node via NFS [19]. Though mature, a central disadvantage of NFS for storing and retrieving high volume data is parallel clients must all contact a single server. A distributed filesystem stores data on multiple nodes and allows clients to access data in parallel from storage nodes, removing this bottleneck. Significant challenges for such a system are scaling to a large number of storage nodes and providing reasonably degraded operation in the face of hardware failure.

In this paper we present StoreTorrent, a distributed filesystem to support high-speed parallel storage and analysis of small record datasets. It provides in-band failure tolerance by replicating data, and scalable performance on many storage nodes. Our system employs the existing NFS service to optimize metadata operations. A significant result of our work is that by efficiently filling and managing a pipeline of in-flight records between the client application and the OS layer, performance for small file I/O operations can be increased several times over existing systems. The presented technique aggregates many small records into larger chunks, reducing the number of system calls made and improving network and disk utilization. Additionally, the metadata aggregates related record information into contiguous data structures that clients retrieve in blocks and process remotely, using indexes to speed queries.

Our initial idea was to logically reverse the popular BitTorrent system to use it for storage rather than file distribution [5,6]. The clean protocol, fault-tolerance, and meta-data management were attractive, and BitTorrent natively segments large files into fixed sized, individually addressable small records for distribution and peer failure resistance. We leverage its protocols and extend them to handle variable sized data records, but eschew its fairness logic and emphasis on data dissemination. In BitTorrent every peer seeks to hold a full copy of the file; in StoreTorrent only a small number of data replicas are made to handle failures. We also retained the tracker class of metadata servers that provide resource discovery to clients via an HTTP web interface.

Conceptually, StoreTorrent overlays a single file namespace on the local filesystems of a commodity cluster connected by a local-area network. Applications use our client library to store and retrieve data. The library employs long-lived, user-level TCP network streams to transport data, and performs redundant writes to guarantee service if some cluster nodes become unavailable, subject to a well defined limit. In addition it provides a quota mechanism and the capability to expand capacity without affecting running applications.

This paper presents the StoreTorrent distributed filesystem, which makes the following contributions:

- Efficient small write performance through application-OS pipelining.
- A lightweight and robust metadata system optimized for small record operations that leverages existing UNIX tools.
- Specific support and significant performance improvement for parallel analysis applications via peer-to-peer communications between storage nodes.

After presenting an application workflow in Section 2, we describe the implementation in Section 3. Section 4 examines failure behavior and presents an analysis of data availability. Section 5 follows with the usage and administration model. Section 6 presents our evaluation of the system. Finally Section 7 describes related work, and Section 8 concludes our paper.

## 2. Application

StoreTorrent holds data for high-speed parallel analysis from a large-volume generator, such as an atomic particle collider, large telescope array, finite-element simulator, or an electronic financial exchange. Let us imagine an instrument that generates data at a constant rate of several hundred megabytes per second, organized in Terabyte datasets, e.g. a single experiment. Let us further assume each experiment can be broken down into many small quanta, on the order of several megabytes, each stored as a named record. Such a dataset will be the subject of analysis, i.e. we seek to store it such that future reads can be done quickly and reliably. A reasonable workflow would store the data from each experiment in a single StoreTorrent file, which contains thousands or millions of records. Finally we will assume we have written a distributed analysis application (e.g. using MapReduce [9]), that can analyze the data for some feature of interest.

Clients running on I/O nodes attached to the generator write data from an experiment. The writer processes use the StoreTorrent client library to send the raw data to a commodity cluster of computers, and multiple I/O nodes may do so in parallel to cluster storage nodes via direct network connections (Figure 1). Optionally the client library sends one copy of each record to F different nodes in the cluster for fault tolerance. Once in a StoreTorrent file, the records exist on the local disks of the cluster, each in a separately named file.

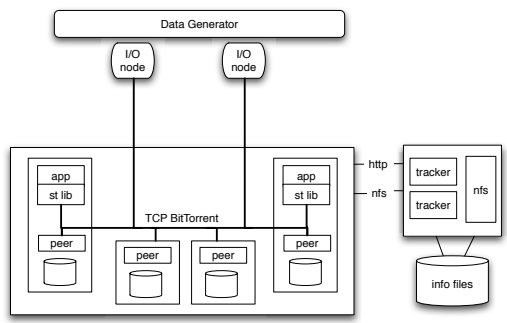

Figure 1) Components. Peers run on cluster storage nodes and save data to local disks. Clients may run on storage nodes or send and receive data from separate machines over the network. Thicker lines are direct TCP connections between clients and peers that carry data using a modified BitTorrent protocol. Thinner lines are HTTP and NFS metadata connections.

Later, the resource manager asks the StoreTorrent metadata system for the nodes that have stored data for the experiment, and runs a process on each one. Each analysis process then requests from the StoreTorrent server running at `localhost` the list of records that happen to have been placed on the same machine, and receives a list of local UNIX file pathnames. It then processes each in turn, reading them directly from local disk, and coordinates with its peers and compute the quantity of interest. The storage servers communicate with each other to hide failures by offering the process a path to a duplicate record if they detect a failed node held the original.

The output of the job can be a small file of final results that are written to an NFS-mounted work directory, or a much larger output of one or more records per input. In the latter case the write bandwidth may be too high for effective NFS service and the analysis task may create a new StoreTorrent file to hold the output.

### 2.1. SPECIFICATION OF SEMANTICS

StoreTorrent supports write-once semantics where new data are not modified unless deleted. This allows simpler consistency protocols that can scale efficiently past a few tens of nodes. In practice write-once systems are quite usable; if an experimenter generates a new data stream they naturally write it to a new filename to be sure old and new data are not mixed. Applications will use a non-filesystem method of synchronization if necessary.

These storage semantics define the basic contract of service to applications [18]. They do not hold if more than one client writes to the same record name simultaneously, in which case the results are undefined, much like when two UNIX processes write to the same file concurrently. All properties below apply to records in a file; the file itself serves as a record namespace.

- **Safe**, *a read not concurrent with a write will return the last value written.*
- **Regular**, *if a read is concurrent with a write, it returns the last value actually written, or the new value.*
- **Atomic**. *A reader can never see the intermediate state during the process of a write, and read/write operations are linearizable.*

Safety is provided by stable storage and failure tolerance (Section 4). Regularity is supported as storage agents write a record to a temporary file and rename it when all data has been received, all record messages are length-prefixed, and the local rename is atomic. The Atomic property is true for new records as the metadata system reveals their existence to readers only after the write is fully completed.

Failures are assumed to be the common crash-recovery type. It is not the goal of StoreTorrent to handle arbitrary failures, and only an initial effort has been made to protect against malicious and rational participants [1].

## 3. Implementation

StoreTorrent has three types of agents: Clients, Peers, and Trackers (Figure 1). Peers are storage servers, named after their BitTorrent counterparts. Clients are applications that use the StoreTorrent library to interact with the system. Trackers create and modify meta-data, monitor peers, and serialize create, write, and delete operations. One and only one tracker exists per file. Clients do not communicate between themselves, and are not assumed to be long-lived. Peers and trackers are long-lived and share a private cryptographic key installed by an out-of-band build system.

Files in StoreTorrent are intended to be read and written by a set of parallel clients running on multiple nodes. Files are subdivided into numerous named chunks of data called *records*, Figure 2. Clients are free to choose a convenient record size, and different sizes may mix in one file, but all should be small enough to easily allocate in memory. It is instructive to imagine a dataset as a stream from the generator instrument, and the writing clients as quantizers of that stream into records. The file is a stream container, and a namespace for records. A StoreTorrent record is the atom of the system: it must be written all at once, may not be split, and cannot be modified unless deleted.

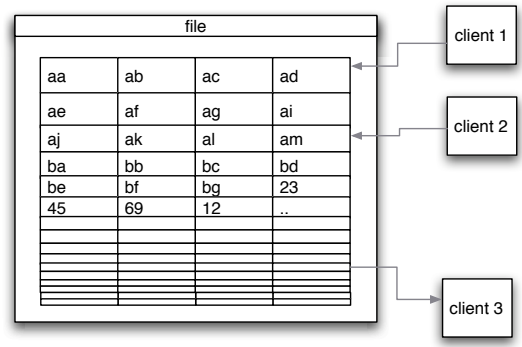

Figure 2) Logical view of a StoreTorrent file. A file acts as a namespace for many named records. Parallel operation involves concurrent access to different records in the file. Client 1 and 2 are writing two records, named "ad" and "am" respectively, while client 3 is reading.

Clients connect with peers directly using a BitTorrent-based protocol over persistent TCP connections, a connection consists of a possibly infinite sequence of messages. All messages are length prefixed, enabling atomic writes of records; if a writer crashes during a PUT, the peer detects the short message and discards the partial data. Our protocol differs from BitTorrent's: there is no choking, no bitfield messages between client and peer, and an entire record is sent in one message, as opposed to 32KB pieces. We introduce several new message types, e.g. to support DELETE and GET_LOCAL operations, and to improve failure reporting.

### 3.1. WRITES: THE PUT OPERATION

Clients receive a list of peer addresses from the tracker on the file create operation and are then free to store records via direct connection to any peer in the list. This allows non-deterministic data placement, which improves failure behavior (Section 4.2). For locality a *blocksize* parameter controls how many consecutive records are stored on one peer before the library chooses another from the peerlist.

When an application has collected record of data, it picks a record name and calls the client library's PUT method with the name and contents (Figure 4). This first contacts the tracker to check quota, then randomly chooses a peer from the peerlist, opening a direct TCP connection to it if one is not already present, and pushes the record using a modified BitTorrent *piece* message format. It also chooses N different peers and pushes the same data again, including in it the copy's *rank*, a unique number assigned to each copy. Only after all peers have responded without error does the client library contact the tracker to *commit* the new metadata indicating the record's existence and peer locations. This scheme, like POSIX, places replication, error handling, and recovery logic with the party most interested in successful storage, the client. It also limits damage from a client crash: any partially written data will at most take up space, but will not pollute the metadata.

On a long periodic timescale, e.g. daily, a scrubber process on every peer removes stale records that were written but never committed in the meta-data due to client failures.

### 3.2. PIPELINING

TCP efficiently manages the network pipeline, and the OS-network interface is carefully optimized in the modern OS. We introduce a technique called application-OS pipelining that leads to improvements in the small write performance of the filesystem.

Pipelining increases the maximum I/O throughput when the application makes a rapid stream of requests. Let us consider writes. In StoreTorrent, the application may queue multiple records to write in a pipeline between itself and the OS, which are sent over the network as a continuous byte stream. The client library need not wait for an ack from each storage server before transmitting the next record, and can naturally coalesce multiple small writes into larger chunks that feed the OS-network pipeline with exactly the amount of data it can accept (Figure 4), saving system calls and improving network utilization. Each request in the pipeline has a unique sequence number, allowing write failures to be tracked individually and out of order.

```
more = True
inflight = f.get_inflight()
while inflight or more:
    if more and len(inflight) < pipeline:
        try:
            f.queue_put( putme.next() )
        except StopIteration:
            more = False
    f.poll()
```

On the peer, this technique allows many contiguous small records to be read at once from the socket buffer and processed efficiently back to back. In Section 6 we show this technique allows small writes and reads to achieve the efficiency of larger chunks.

### 3.3. PEER MONITORING AND SCHEDULING

The scheduling of peers to a file is done at create time. The tracker requires the client to provide the estimated total size of the file, and schedules peers to the new file based on that figure, how many peers are available, and how full they are. This task must be done carefully to avoid hot spots of overly full peers [36].

The tracker gathers peer information including bytes of local space used and available using BitTorrent

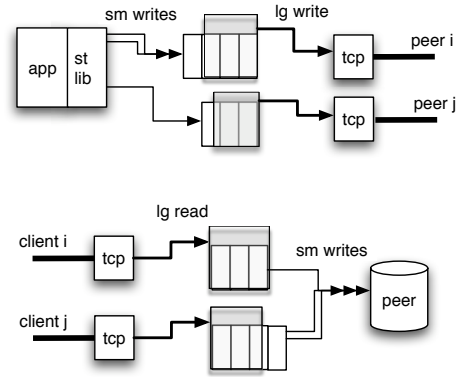

Figure 3) Pipelining for small write performance. On top, a writing client queues multiple outstanding writes to the pipelines. A single write call, e.g. coordinated by poll(), fills the TCP buffer to each of two peers. Below, a peer makes large reads from two client connections then decodes them into component records and stores them to disk in new files. The local filesystem will allocate new blocks contiguously.

*announce* messages, which are pushed periodically from peers. This soft-state approach for obtaining global system knowledge has good resilience to failures and naturally handles peer arrivals and departures. The tracker webserver's root URL provides a summary of these statistics to browsers and a Python monitoring library, including the number of alive/failed peers, number of files created, the filesystem total and available size.

### 3.4. PEER STORE

Peers employ the underlying local UNIX filesystem on a node to store data, as they efficiently use storage hardware and rarely fail or corrupt data even under extreme conditions [25]. This additionally ensures a peer crash during a write operation has well-known semantics; the state of the data is the same as any interrupted process writing to the local UNIX filesystem. This design choice frees us from the concern of maintaining internally consistent storage data structures, a problem known to be hard [38].

The protocol is agnostic to any reasonable storage scheme implemented by peers. Ours simply spread records evenly in files and directories under a local base directory. The use of a single file per record, as opposed to employing fixed sized chunks, simplifies the parallel operation of the system and the periodic scrubber that cleans up after client failure. It also ensures no record is split between chunks. In addition to record data, the PUT message includes a CRC which the peer keeps in a separate file and returns unaltered on GET. The client library verifies the checksum against that of the received record contents to provide end-to-end protection against silent read errors from peer disks.

### 3.5. METADATA

The metadata points reading clients to the peers holding a given record, and supports enumerating all records in a file. Like BitTorrent, we consolidate all record metadata for a given file in a single data structure. This allows clients to make efficient block reads for record metadata, reducing the number of network round trips during operations and the load on the metadata server.

Metadata files are stored at a central location and updated by a Tracker agent (Figure 1). A second copy of metadata is stored locally by each peer, which contains only a per-file list of records held. The latter is used for parallel analysis GET_LOCAL, the former by the GET, both described later in this section.

Files in StoreTorrent are named by a familiar UNIX pathname, symmetrical to the name of an *infofile* that stores metadata for that file. The info file holds an SQLite file-based database [14]. For example the StoreTorrent file named "/foo/bar/baz" has an info file visible to clients at /foo/bar/baz. The elements "foo" and "bar" are regular UNIX directories. Symbolic links and file permissions on infofiles have the expected semantics, and UNIX tools such as `find` and `tar` can be used normally.

Info files are visible on the network via NFS using the fileserver present in the environment. Clients mount the infofile volume read-only, while the file tracker runs on the same machine as the NFS server and has read-write access. A tracker is only involved in mutating operations: create, commit, and delete. The interface between the tracker and NFS is the read-write filesystem containing infofiles on the tracker machine and the ACID semantics of SQLite. This separation is motivated by the observation that in a write-once system supporting analysis, read load will be greater than write load, and the optimized in-kernel NFS server is better suited where practical than a user-mode agent. It is made possible by the fact that while the SQLite library requires local filesystem access to infofiles for writes, it safely supports NFS access for reads.

We further reduce read-side metadata load by pushing SQL processing to the client machines, rather than employing a centralized RDBMS. The NFS server provides only data blocks to SQLite libraries that run on client machines, which perform the bulk of processing tasks such as record enumeration, search, and selection. SQLite maintains indexes on record names to improve processing performance for strided or name-based read patterns.

Clients contact a tracker at a well known address, e.g. as specified in a site-wide configuration file, via an

```
f = st.create(filename,ft_class,est_size)
f = st.open(filename)

f.put(contents, name)
f.queue_put(contents, name)
contents = f.get(name)
f.queue_get(name, callback_on_done)

iter = f.get_local()
for name, fn in iter:
    open(fn).read()

f.delete(name)
f.close()
```

Figure 4) API. The StoreTorrent client library is designed for parallel operation and provides an interface including the primitives shown above. The "name" argument is a record name string. The library provides both blocking and non-blocking (queue_*) versions. Some details such as open mode and app-specified file variables and record attributes are omitted for clarity.

HTTP/HTTPS interface, similar to BitTorrent. Two optimizations help reduce the tracker load. First, the client library groups records together before initiating tracker operations, allowing the tracker to update the metadata of many records at once [12]. Second, the master tracker running at the well-known address may spawn per-file trackers to handle metadata load for one or more files. Clients are redirected to the appropriate tracker on open.

If the tracker daemons fail or are unacceptably delayed, clients continue to have read-only access to data while the NFS server is operational. Two servers sharing a network block device will survive common hardware failures and is a common high-availability configuration [40]. The NFS server and trackers would run on one or both server heads.

Permission to read and write records for the StoreTorrent file on peers is regulated by the permission bits of the info file and its parent directories. This access control is safer than NFS since the tracker demands strong certificates from clients to prove their identity (our implementation uses Munge [11]), and only the tracker has read-write access to info files. Peers refuse to store data for clients unless presented with a per-file signed certificate from the tracker, which clients obtain on create. Peers and the tracker share a private cryptographic key for this purpose. Therefore while read access may be spoofed due to the insecure NFS protocol, write access cannot.

The metadata in the centralized info tree is shadowed by metadata held by storage peers themselves. Indeed critical metadata state may be (slowly) reconstructed by contacting all peers in the system, a useful property in the case of catastrophic failure.

## 3.6. READS: THE GET OPERATION

The first read method is a GET operation that connects to one of the peers listed for the record in the meta-data, and directly pulls its contents over the network. The client library takes a filename as input and issues an SQL select for the desired record(s) to the infofile found there. Among the metadata returned is the peer index, which is used to lookup its IP address and port. A new TCP connection is made to the peer if one is not already open, and the whole record is requested. The server replies with the contents in a *piece* message of the same format used for PUT calls, including the given CRC. No contact with the tracker is required.

In the common case several peers are listed, and the client library requests the record from one at a time. This operation succeeds if at least one peer holding it is alive. Clients do not request the record from all peers in parallel to avoid overloading the client's incoming network pipe, a problem known as *Incast* that can cause a 10x collapse of useful bandwidth [23].

## 3.7. READS: THE GET_LOCAL OPERATION

Parallel analysis jobs use the second read method, GET_LOCAL, which reveals locally-stored records to clients. The task is to enumerate all file records, despite peer failures, to a coordinated set of analysis processes run directly on the storage nodes. These applications are agnostic to *which* data is read, as long as in aggregate the reads are disjoint and complete, i.e. each record in the file has been seen by some process. This relies on the commutative and associative properties of the analysis, such that the order of reads does not affect the computation. All records in the file are revealed in the form of a list of *('record name', 'path')* tuples. If the analysis requires only a subset of records, it iterates through the list to determine the ones of interest by name.

Peers communicate among themselves to discover the set of available record replicas, without tracker contact or access to the central metadata store. Following are invariants the system maintains during the operation.

1. *A record is revealed to at most one client*. This ensures each record is processed only once, regardless how many duplicates exist in the system.
2. *Only one record exists at a given copy rank*. If this invariant breaks the operation will fail, identifying the offending peers.
3. *A peer holds only one copy of a record*. This safety property must hold to get any benefit from duplication.

At job initiation the resource manager starts one or more analysis processes on each storage node in the file's peerlist. Each process opens the file and calls GET_LOCAL, which contacts its local peer and provides it the peerlist. Peers in the list contact each other with direct TCP connections and perform a ring-based *AllReduce* operation [32] to exchange indexes of records each holds and at what copy rank. The collective streams each peer's metadata around the ring in N=len(peerlist) rounds, allowing each peer to gain global knowledge of all record copies present. This transaction is initiated by the first peer to be contacted by a client, and can complete among peers without further client involvement.

Peers build a list of record pathnames, valid on the local machine, to return to their local client. These are later accessed directly by the client using normal open/read/close, with no further aid from the peer. We choose that lower copy ranks take precedence; peers always list all copy rank 0 records they hold for the file. Peers list a record copy rank 1 if no other alive peer holds the rank 0 copy. For a general number of record copies, a peer lists a local record copy rank F if and only if no peers hold copies of that record in ranks (0,..,F-1). After completing the algorithm, peers cache the result for 60s for late-arriving clients.

With this support, the filesystem can greatly aid application performance by exposing its location information, enabling the application to prefer local data to remote. In the expected case the execution will transfer *no data* over the network. This method is not new. Hadoop employs this strategy via a central metadata server for its MapReduce engine [41], but not for general HDFS clients. The NUFA facility of Gluster [39] prefers local storage over remote. StoreTorrent is the first filesystem we are aware of, however, to use copy-location knowledge to avoid transferring data even in a degraded state.

## 4. Failure Behavior

In this section we examine the system's reaction to failures during operations, and provide an analysis of unavailable records due to excessive storage node failures. In related papers this section is sometimes called *High Availability*.

## 4.1. DATA REDUNDANCY

Failures must be expected for any electrical device, and have been specifically studied for clusters of commodity components [24,28]. StoreTorrent however does not force a particular redundancy scheme on applications. It supports a *fault tolerant class*, chosen per file at create time, which s responsible for replicating the data as the system's defense against failures. The method is not expressed as an integer describing a number of storage node failures but instead as a class name, to allow flexibility in specifying new methods, e.g. *raid5* and *x2* both handle 1 failure. The client library wholly implements the class, its operation is invisible to peers, who simply see additional write requests.

As any redundant data must be stored normally, fault tolerant classes reduce the usable capacity of the filesystem. Some schemes, however, require less than a full replica to tolerate a failure. An early version of StoreTorrent employed an XOR based *raid5* class that split each record into S pieces plus a parity piece, and stored each on a separate node of a static *peer stripe*. This had the advantage of fault tolerance with the fractional capacity overhead of 1/S, but limited the file size and read performance. The largest practical stripe size was found to be approximately 10 peers, due to Incast effects that overloaded reader's network connections. While Panasas uses a 2-level raid scheme to overcome this limitation [36], any parity technique leaves only partial records on storage nodes, precluding the use of the GET_LOCAL method.

The *x2* class, in contrast, does not split records, but stores two copies, unmodified, on different nodes. This class is designed for thousands of peers, and supports GET_LOCAL access of records. Classes *xF* for positive integer values of F are available to handle F-1 storage node failures, including an x1 class that makes no copies for applications that do not need fault tolerance.

### 4.2. ANALYSIS OF DATALOSS

Let $t$ be the number of storage node failures a data redundancy method tolerates. What is the affect of additional failures? In a common scheme, a whole storage node is replicated, in a Raid1-style stripe. If the data is stored evenly over N nodes, and $t$ failures have been sustained in the raid stripe, an additional node failure in the stripe causes $O(1/N)$ of the data to become unavailable until the node is repaired. In StoreTorrent, however, an additional failure results in only $O(1/(N^2))$ of the data to become unavailable because of the non-deterministic data placement.

Let us assume 100 storage nodes in the filesystem, where 1000 files have been written, each with 100,000 records. Files used the x2 class, so each record was written to two distinct nodes. These files occupy 20TB in the filesystem, with 40TB stored including duplicates. The peerlist for each file encompasses all 100 nodes, so each record has an equal chance of landing on any node.

No data is lost if one peer fails. We will analyze how many of the 10 million records become unavailable if more failures occur. At one additional failure, i.e. 2% of nodes, only 0.01% of records will be unavailable. These are the ones that have both copies, x and y, on the two failed nodes. The expected number of inaccessible records is given by:

P(x in C, y in C) = 1/100 * 1/99 = 1/9900

= .0101% records unavailable

where C is the set of failed nodes. Note the PUT ensures x≠y. More generally, in files using F record copies the first F failures cause no interruption. If N is the total number of storage nodes, the probability of an unavailable record after G additional node failures is:

P(all F copies in C) = G/N*G/(N-1)*...G/(N-F)

For F << N this approaches $(G/N)^F$

Our result does not require the failures to be independent, e.g. it is unaffected by nodes failing in the same rack or two "twin" nodes loosing a shared power supply, etc.

In a Raid5-style stripe of size S, $t$ < S-1, i.e. each stripe holds more data than in Raid1, making its failure more costly. This is generally ameliorated by rapid and automatic stripe rebuild to other storage nodes, but a stripe-correlated set of failures could overcome this rebuild before it is finished.

### 4.3. PEER FAULTS

A failure may be either intentional or unintentional: it may occur due to reasons including a software error, an administrator's signal, a hardware failure, or a network switch going offline. We define faults in a simple way: *the peer could not perform the task requested of it*. The clients use a coarse socket timeout to detect hangs, and peers send explicit failure messages when possible. The library discriminates between the two, as an explicit failure may be traced to a specific record identified by a sequence number, while a timeout or socket error (e.g. a port unreachable ICMP reply) is determined to affect all records from that peer. The tracker declares faulty any peer that skips several announce messages, but clients may also explicitly notify the tracker of peer failures they detect.

The StoreTorrent client library transparently handles failover on observed peer faults. There are two cases: a peer fails well before the operation starts, called an *initial failure*, and the peer fails just before or during the operation, a *dynamic failure*. For initial failures the tracker includes a list of failed peers in the open response, and clients then avoid the suspect peers. For dynamic failures let us consider the x2 class. If a peer fails during a PUT, the client simply chooses a new peer from the list; the operation will succeed if there are at least two functioning peers present. For GET, the client will request the record from the first peer listed, if that fails, the second, and if not successful, will fail the request. For DELETE, the client first asks both peers to delete, then notifies the tracker. To cover dynamic failures that occurred during delete, a periodic scrubber process run locally on peers inventories all records, deleting those not listed in their corresponding info file.

### 4.4. WRITER CRASH AND REBALANCE

What if the writer crashes after sending pieces to peers but before calling commit? Like UNIX writes, StoreTorrent expects the application to retry. Since a new PUT is atomic, partial writes have no ill effect to the data regardless of when the crash occurred. However a delete may be necessary to reset the two-phase-commit metadata state of the tracker, which it uses to enforce quotas. If the peer detects a client failure via a timeout or unexpected socket close, it notes an error in its logfile, removes any unfinished record tempfiles, and cleans up the connection.

To maintain safety, a record cannot be modified before deletion. If such semantics are desired, a more complex algorithm is needed to ensure updates are protected from temporarily unavailable nodes, which later return and unknowingly present stale data. It is therefore temping to simplify the system by strictly prohibiting record rewrite. However StoreTorrent supports rewrites for copy rebalancing in the case of a failed but unrecoverable peer. The client library ensures that on a record rewrite all alive peers remain as record copy holders, and only failed peers are replaced. This allows a simple *degraded-read*-then-*write* sequence to function as a rebalance operation to restore the F record copies. A higher level monitoring service currently initiates this rebalance as necessary.

### 4.5. GET_LOCAL

If a rebalance operation has been performed after peer failure, the failed peer's should not rejoin the service without having its local store deleted. If this is not done the protocol may encounter two peers holding the same record at the same rank during the all-peers collective, and will return an error to protect invariant 2. Practically this should not be an issue as rebalance accompanies manual actions to address a failed storage node, which for StoreTorrent will include wiping its disks.

Peers expect a fully alive peerlist as input. Using timeouts parameterized on the peerlist size, they will notify the client library of failure, which then removes the offending peer and restarts the operation. If more peers have failed than is tolerated by the data redundancy scheme of the file, the client library notifies the application.

## 5. Usage and Management

As pointed out by system administrators [7] daemons implementing a DFS should respond immediately when given a signal. StoreTorrent adheres to this rule: daemons tear down open connections and stop immediately when killed. Trackers and peers make no distinction between normal and abnormal termination. Similarly when a client application receives a stop signal from a user's terminal, it stops promptly. As there is no kernel code, an error should not crash the OS or hang the process in uninterruptable disk wait state.

The tracker stores all hard state, defined as that which cannot be recovered passively during normal operations, in the infofile databases to ensure unexpected restarts do not leave the system in an inconsistent state.

StoreTorrent can expand capacity by adding any number of new cluster nodes, without taking the system offline or disrupting active clients. The new peers are immediately available to accept new files after their first announce. StoreTorrent inherits this behavior from BitTorrent, where new peers are added and removed often during operations. No effort is made to automatically rebalance existing files after expansion; this process is currently implemented by a higher layer process that copies a file to a newly created file backed by new peers, and then deletes the original. Removing a storage node is treated like a normal failure.

Finally all storage nodes are symmetrical: there is no need for an array of special "NAS head" machines to interface between clients and data storage.

## 6. Evaluation

We tested the performance of StoreTorrent with a typical workflow for its intended domain, similar to that described in Section 2; we write a large amount of data to a file in parallel, then read it back from a simulated parallel analysis task. We then ported a production MapReduce-style analysis job to use the StoreTorrent client library and measured its read performance.

### 6.1. IMPLEMENTATION

Our current implementation uses a Python client library and C++ peer, and no kernel-space code other than the widely available kernel NFS server. The implementation is small: the client library is 4300 lines of Python, the storage peer is 3000 lines of boost/C++. The test code for correctness under fault conditions occupies more than 3100 lines of Python. All agents use an event-based programming model for CPU cache friendliness and performance with many concurrent connections [17], and use the Linux `sendfile` and `epoll` system calls where appropriate.

### 6.2. METHOD

We performed our evaluation on 70 nodes of a commodity cluster running Linux 2.6.26 x86_64 using the Rocks Cluster Distribution version 4.1 [22]. The nodes were approximately homogenous, each with 8 Intel 5410 CPU cores running at 2.33GHz and four 500GB, 7200 rpm SATA disks, arranged in a software raid5 array. Data was stored on an ext3 filesystem made on this array, which achieved a maximum performance of 57 MB/s for block writes and 117 MB/s for block reads using Bonnie++. The network consisted of copper Gigabit Ethernet from nodes, and switches connected with 10 Gigabit Ethernet to provide full bisection bandwidth. The StoreTorrent client library used Python version 2.5.1 and peers were compiled with gcc version 4.2. Gluster was evaluated on a subsequent configuration of the cluster, with the four disks arranged as a Raid0 array.

StoreTorrent clients were configured with a pipeline depth of 10 records, and made tracker calls with a group commit size of 40 records. Each trial used one file with a peerlist that included all 70 nodes.

We evaluated two existing open source distributed filesystems for comparison, both of which use local filesystems on cluster nodes for storage. The first is PVFS, version 2.7.1 [3]. We chose PVFS due to its maturity and explicit goal of high performance. This filesystem also uses cluster nodes as storage servers, and clients communicate directly with them to store and retrieve data. PVFS employs one or more metadata servers that track the location of files, and uses a Linux kernel module to provide a POSIX interface to clients. In its native configuration, PVFS does not handle a node failure without data loss. However PVFS provides fault tolerance when a storage unit is in a two-node primary-spare server configuration with both connected to a network block device, such that the spare assumes load on failure [40]. This has the advantage that data is not replicated, but its extra cost and hardware complexity precluded our testing this configuration. A single meta-data server is used as it provided higher performance in our tests than running each PVFS server as both a storage and meta-data server.

The second system is GlusterFS [39], using version 2.0.4 and Linux 2.6.30 on the same hardware. We chose this distributed filesystem due to its high performance, scalability, and support for a local-storage-preferred mode called NUFA, similar to StoreTorrent's GET_LOCAL. Gluster uses the FUSE user-level filesystem to provide a POSIX interface, and employs no central metadata server. Instead, the DHT facility of the library stores metadata in ext3 extended attributes in directories of the local filesystem, which are replicated to all nodes. These attributes include a map of hash ranges that point clients to the server that stores a file, based on that filename's hash. Gluster supports fault tolerance, inline to the protocol, via whole node replication. The replication factor is static and applies to all files; its parameters are set by an input file at system startup. We configured Gluster in NUFA+ REPLICATE mode that duplicated one node's data to a partner node Raid1-style. The storage daemons used 8 I/O threads, one per CPU core, as we measured the highest performance with that setting.

### 6.2.1. PARALLEL WRITES

This experiment examined the bandwidth available to a set of clients writing data to the filesystem. In all cases writers ran on the same 70 nodes that provided storage, and the written data was present in writing processes memory before the timings were started. The same client code was used for writes to each filesystem, in different modes: it used the client library for StoreTorrent and POSIX open, write, and close for PVFS and Gluster.

To find a lower bound on performance we took steps to simulate a loaded environment for the filesystems, e.g. when a storage node has exhausted its filesystem cache and every write leads to a cache eviction and disk commit. To this end we dropped the filesystem dentry, inode, and buffer memory caches before every run. StoreTorrent peers first called fsync on a newly written record, and then invoked `posix_fadvise(DONTNEED)` to immediately discard it from the FS cache. For PVFS we only forced the fsync with the TroveDataSync=yes setting, but took no steps to flush the filesystem cache during a trial. For Gluster we disabled all available performance caches in the configuration file.

For the small write test, each client starts in parallel and generates a number of 2MB buffers in memory, using random binary data, then starts a timer. All clients open the file, then PUT a record with the contents from a buffer in the generated set until 32 GB per node has been written, or 1146880 records over all 70 nodes. A trial employs a single StoreTorrent file; for POSIX filesystems this is represented by a directory, and the driver code writes a record as a file in that directory. The number of clients (P) was increased from 1-560. The vertical orange lines highlight P=70 for one writer per node, and P=560 for one writer per CPU core. When there is more than one client per node, each writes a fraction of the 32GB.

Gluster provides efficient write performance with large 64MB writes, achieving 48.4 MB/s/node at 8 client writers per node. With small 2MB writes, however, this falls by a factor of 10 to the value of 4.1 MB/s/node shown in Figure 5. StoreTorrent reached 22.6 MB/s/node for the same small writes. This is due to application-OS pipelining (Section 3.2). As PVFS does not replicate data for fault tolerance, it stores half the data of the StoreTorrent and Gluster trials. It performs well for small writes, however, from 1-4 writers per node. At 8 the meta-data server became a bottleneck.

Each datapoint is the average of at least three runs with identical parameters. During analysis of these results we found that some runs experienced a 2-10x collapse of observed bandwidth. This is due to the experiment operating at the edge overloading the incoming 1Gbps Ethernet link to the storage peers, or in the case of readers, the client machine. As described by Phanishayee et al [23] this *Incast* problem is difficult to correct at the node but larger network switch buffers or faster networks may help.

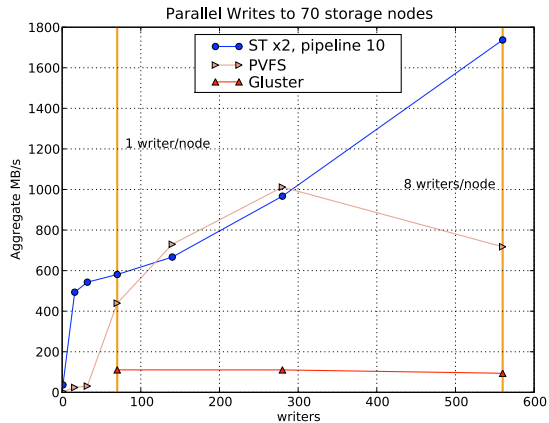

Figure 5) Writes. 70 to 560 processes running on storage nodes write 2MB records to the filesystem in parallel. 70 writers is one per node and 560 is one per cpu core. All data is flushed to disk immediately where possible, and filesystem caches were dropped prior to each run. ST and Gluster send twice as much data as PVFS to provide in-band fault tolerance. Gluster is configured in NUFA mode and no less than one client per node was run.

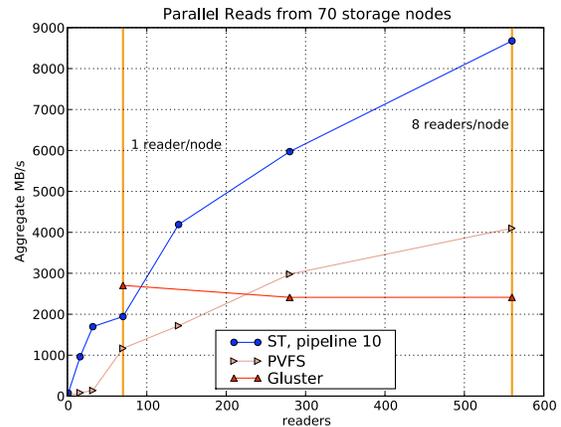

Figure 6) Reads. Parallel clients read a file with 1.1 million records over the network. Both ST and PVFS send every byte over the network, but Gluster favors local data if available (using its NUFA+replicate translators). ST gives no bias to record location in this test, no files are read directly by clients. The filesystem's cache was cleared before each datapoint was gathered.

### 6.2.2. PARALLEL READS

In this experiment 2MB records are read over the network by 1-560 parallel client processes. This simulates dataset access from non-local processes such as workstations, or analysis tasks run on a cluster that do not have the commutative and associative properties for embarrassingly parallel operation. No local file-access is performed, all bytes are sent over the network.

The same client driver is again used for all filesystems, and a file from the previous experiment is read. All clients start a timer then open the file in parallel and choose a disjoint slice of available records to read, decomposing the range by their numerical rank. Each client reads a record at a time into a Python string, and then discards it, until all records in the slice have been processed. Each reader computes its own read bandwidth, the aggregate of which is shown in Figure 6. Again each datapoint is the average of three trials, and the filesystem memory caches were dropped before each run to force reads from slow physical storage and to determine a lower performance bound.

StoreTorrent's small read performance scales well, reaching over 100 MB/s/node as some records are read over the loopback network device. Gluster also provides comparable performance for large 64MB reads at 63.2 MB/s/node. Its 2MB read bandwidth of 27.1 MB/s/node is slower by a factor of 2.3. The PVFS read performance scales throughout the range.

During the write and read experiments StoreTorrent clients contacted a central metadata agent: for writes the file's tracker, for reads the NFS server. We found a group-commit size of 40 allowed the tracker to make efficient batch updates to the infofile; larger values did not reduce its load during trials significantly. During writes load on our python Tracker occupied less than %5 of the machine's capacity at 560 clients. During reads the NFS server load was less than 2% capacity at 560 clients.

### 6.2.3. PARALLEL READ_LOCAL

In this experiment we launch 1-560 clients on the storage nodes, start a timer, then request all local records for a file from the peer at `localhost`, and read the resulting list sequentially. When more than one client ran per node, the library decomposed the list among them such that each record was read once. The experiment measured the maximum read rate achievable for a 40 GB file and a 1.5 TB file, each composed of 2 MB records. No filesystem memory caches were altered between write and read, and the cluster was otherwise idle during each trial. Clients stop their bandwidth timer after they have read the last allotted record, and like the previous test read each into a Python string. The smaller file fits wholly in the filesystem caches of the storage nodes, the 1.5 TB file does not.

The results are shown in Figure 7. The rates include the time to perform the AllReduce collective between peers; this phase never exceeded 2% of the measured read time. Neither POSIX filesystems provided similar functionality and so are omitted from the figure. The 1.5

TB file read shows the disks are nearly saturated by read requests at 4 clients per node.

*6.2.4. AN ANALYSIS WORKLOAD*

In the final experiment we measured the maximum performance of the read phase of a parallel analysis job. We ported a MapReduce-style analysis framework [33] to use the StoreTorrent GET_LOCAL call, requiring only a few new lines of code. StoreTorrent peers provide record locations to map processes running on each node, which read and pass their contents to the map function. The pivot and reduce steps of the analysis were unaltered, and results were written to an NFS working directory. For the experiment we used an existing analysis code with the ported MapReduce library and ran it on a dataset from a finite-element simulation trace generated from an HPC cluster.

This experiment sought to characterize the maximum performance for a representative analysis task performed immediately after data generation. We wrote the 55 GB dataset of 30,000 1.9MB records into a 70-node StoreTorrent file while the cluster was otherwise idle, then immediately launched the analysis task on all 70 nodes. We observed the map processes sourced and parsed data in to usable Python sequences at the rate of 900 MB/s/node. The data set fit into the aggregate filesystem memory cache; the rate reflects speedup from cache hits, and this figure represents maximum achievable rate for the analysis task on this cluster.

We note that if the storage system is used by one parallel analysis task at a time, each peer's filesystem cache will be aggregated and at the disposal of the job. As shown, the application can expect a super-linear speedup of read performance with increasing number of nodes if the file is warm in cache, i.e. if it has just been written or the analysis is performed multiple times in succession.

# 7. Related Work

Many recent efforts, both commercial and academic, have targeted large scale distributed storage. Most seek to provide speedup for unaltered legacy applications using a kernel module to provide a POSIX filesystem interface.

Commercial offerings include Panasas [31,36] and Ibrix [30]. Panasas employs commodity servers with battery backup for storage, and uses client-direct Raid5 encoding for fault tolerance. In order to support large files but keep the raid stripe size small to mitigate Incast, the system uses a 2-level raid scheme that assigns different stripes of servers (parity groups) to each GB-sized file chunk. Ibrix stores chunks of data called segments on storage servers, and stores the segmentID in a file's inode. The filesystem broadcasts a

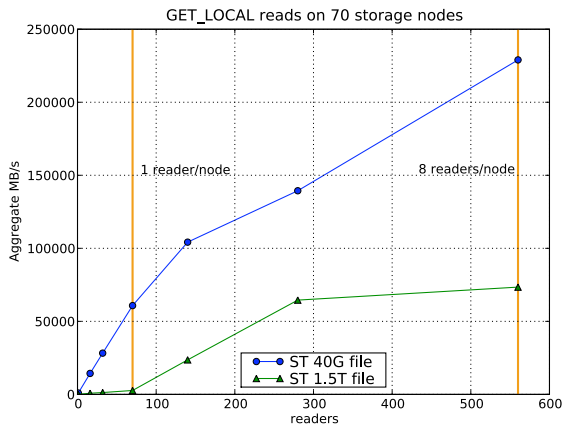

Figure 7) Read local. One to 560 ST clients request all local records and read them sequentially. No data is read over the network, the only inter-node communication is the peer-to-peer collective, which in all cases takes < 2% of the overall time. The 40G file fits wholly in filesystem cache of the nodes, the 1.5T file does not.

(segmentID, server address) map to all clients that guide reads to the correct server.

pNFS [13] is a version of NFS that separates data from metadata traffic, to allow clients to access data in parallel from multiple servers. Recent studies use PVFS2 as the backend block engine, and show scalability with tests that use 8 or fewer storage nodes [4,15,16].

Recent work addresses high performance writes for the purpose of supporting large cluster checkpointing load. These act as a cache for a parallel filesystem and do not have failure tolerance or read support. The system from LLNL [20] uses a *node-local* strategy to favor disks local to the write process when possible, to achieve very high performance. Zest [21] sends all writes over the network, employs non-deterministic data location, and also achieves high write efficiency.

Warren suggested a high performance storage architecture [34] that provides a SAN-like shared block device at low cost. It uses Linux software raid to aggregate iSCSI block devices running on nodes of a cluster. This system is attractive but must do a time-consuming re-sync operation on failure, and is subject to the Incast problem with many storage nodes. The commercial Exanodes [29] provides a similar block device but gracefully handles re-replication on failure. Both rely on existing filesystems such as ext3 for single clients, or a clustered file system such as RedHat's GFS, IBM's GPFS or Oracle's OCFS2 to support parallel clients.

Lustre [2] is a well-known free distributed object storage system. Meta-data is handled by a single server that may failover to a backup, and data is stored on reliable nodes. Ceph [35] is a recent open source system with a similar design to Lustre but balances metadata load over multiple servers.

Hadoop [41] is an open-source Java implementation of the original MapReduce framework [9]. It provides a non-POSIX filesystem, HDFS, which stores files in large chunks on disks of a commodity cluster, with in-band fault tolerance via replication. The filesystem uses a centralized metadata server and employs data-location knowledge to prefer local data during map tasks. Cloudera has shown Hadoop/HDFS has performance issues with small file I/O [37].

Storage systems for the wide area [8,27] emerged in the last few years, which use an efficient distributed hash table (DHT) algorithm to coordinate service. These systems target very large sets of loosely coupled and unreliable nodes, requiring highly redundant data and careful request routing. In contrast StoreTorrent assumes dedicated, tightly coupled storage servers and uses simple protocols to achieve high performance. Finally Pond [26] is a cluster storage system that employs a DHT for metadata service, and explored the use of forward-error-correcting for fault tolerance with fractional capacity overhead. They determined that while attractive, such codes are too slow for first-line storage.

Local filesystems have also been optimized for streaming writes and analysis. StreamFS [10] supports efficient small writes to disk for packet logging via careful handling of the disk head and optimized metadata. If modified to support general reads it may well serve as local storage for an analysis tasks on a distributed filesystem.

# 8. Conclusion

In this paper we presented the design and evaluation of StoreTorrent, a distributed filesystem for storing Terabyte datasets on the local disks of a commodity cluster. We described the dual requirements of fault-tolerant data access and high performance reads, and showed how our implementation achieves these goals. We described a local read method to support parallel analysis tasks, which is novel in its use of data copy location information to provide un-degraded performance in the face of failures. We presented an application-OS pipelining technique and showed it improves the efficiency of small writes and reads by a factor of 1-10. We showed how the system's non-deterministic data placement during writes has the benefit of exponentially reducing the amount of unavailable data caused by excessive storage node failures.

The lightweight and robust metadata system leverages the mature NFS service and supports many existing UNIX filesystem tools for administering the system. We evaluated StoreTorrent on a simple inexpensive cluster using 70 nodes and showed scalable, efficient performance for parallel write, read, and local-read operations needed to store raw output from large-scale data generators, and analyze it to produce useful results.

## 8.1. FUTURE WORK

While our evaluation cluster had 70 nodes, our target was 1000. Since the tracker can limit the peerlist size for each new file, given enough offered load over different files the system should distribute it correctly over 1000 nodes. However we hope in a future evaluation to explore peerlist size limits for a single file, which likely will require techniques to limit the number of open TCP connections from a client.

### *8.1.1. SECURITY*

While the system was built with security in mind, much of the implementation is left for future work. While file creation and storage participation is protected by strong userid certificates, the specific transitions of the protocol are not, opening the way for rational clients to obtain more benefits from service than intended. We observe that good security protects against innocently malfunctioning client implementations as well as rational or malicious attacks. For example a replay "attack" could come from a buggy client that used an old certificate to inject bogus records into a file, consuming quota and confusing readers.

### *8.1.2. POSIX INTERFACE*

It may be possible to model the StoreTorrent interface in POSIX with a directory per file, and each record represented by a file in that directory. Each record would be written with posix calls `open(file_dir/record_name); write (contents); close()`. A special `info/` directory could hold peer address information. However a pure POSIX interface that can be used by unmodified applications may not be possible.

## 8.2. ACKNOWLEDGEMENTS

Place holder.